\documentclass[aps,twocolumn,floatfix]{revtex4}%
\usepackage{amsfonts}
\usepackage{amsmath}
\usepackage{amssymb}
\usepackage{graphicx}%
\renewcommand{\H}{\mathop{\mathrm{H}}\nolimits}
\begin{document}
\preprint{ }
\title{Time-dependent transport via a quantum shuttle}
\author{M. Tahir$^{\ast}$ and A. MacKinnon}
\affiliation{Department of Physics, The Blackett Laboratory, Imperial College London, South
Kensington Campus, London SW7 2AZ, United Kingdom}

\begin{abstract}
We present a theoretical study of \ time-dependent transport via a quantum
shuttle within the non-equilibrium Green's function technique. An arbitrary
voltage is applied to the tunnel junction and electrons in the leads are
considered to be at zero temperature. The transient and the steady state
behavior of the system is considered here in order to explore the quantum
dynamics of the shuttle device as a function of time and applied bias. The
properties of the phonon distribution of the oscillating dot coupled to the
electrons are investigated using a non-perturbative approach. We derive a
relation for the oscillator momentum charge density correlation function which
is an interesting physical example for the visualization of shuttling
phenomenon. We consider the crossover between the tunneling and shuttling
regimes for different values of the key parameters as a function of applied
bias and time. We also consider the energy transferred from the electrons to
the oscillating dot as a function of time. This will provide useful insight
for the design of experiments aimed at studying the quantum behavior of a
shuttling device.

\end{abstract}
\startpage{01}
\endpage{02}
\maketitle

\section{Introduction}

In recent years, there has been a great interest in the study of transport
phenomena through nanoelectromechanical systems (NEMS)\cite{1,2,3,4}\ both at
an applied and at fundamental level. This attention has been focused due to
the fact that novel transport regimes arise due to strong interplay between
electrical and mechanical degrees of freedom, which promise both the concept
and realization of these systems as a new generation of quantum
electromechanical devices. A large number of new experimental techniques have
been developed to fabricate and perform experiments with NEMS in the quantum
regime; Examples of high-frequency mechanical nano-structures that have been
produced are nano-scale resonators\cite{5,6}, semiconductor quantum dots and
single molecules\cite{7,8,9,10,11}, cantilevers\cite{12,13}, vibrating crystal
beams\cite{5} and more recently graphene sheets\cite{14} and carbon
nanotubes\cite{15,16}. Anticipated quantum effects\cite{5,6,12,13,17} include
observation of single phonons as well as ground state effects such as zero
point displacement, ultra small mass sensing, measurement of extremely weak
forces and suppression of scattering in coupled electromechanical systems. The
strong interplay between a phonon mode and electrons is crucial for observing
the quantum effects in the transport of charge through electromechanical systems.

A special kind of electromechanical system where the interplay between the
electrical and mechanical degrees of freedom drastically changes the transport
properties is a shuttling device\cite{7,18,19}. The characteristic component
that gives the name to these devices is an oscillating quantum dot of
nanometer size that transfers electrons one-by-one between a source and a
drain lead. A shuttling device\ is a particular example of a NEMS device, an
exciting generation of electronic devices such as: sensitive mechanical charge
detector, logic gates, sensors, and memories. Beside these applications, a
charge shuttle may increase the understanding of quantum transport properties
of NEMS and related quantum phenomena in nanoscopic systems. In an original
proposal for a charge shuttle by Gorelik, et al\cite{20}., the investigation
of the quantization of charge in the strong Coulomb blockade regime was
discussed. This system consists of a movable, micro-size metallic grain
spatially confined in a harmonic potential between source and drain leads. In
the strong Coulomb blockade regime only a single excess electron at a time is
allowed to occupy the grain. When a sufficiently high bias is applied between
the leads, a single electron may tunnel onto the grain, and due to the
electrostatic field between the leads, the charged grain is driven towards the
drain, where the electron tunnels off. Due to the harmonic potential the
uncharged grain is forced back towards the source and the process is repeated.
A mesoscopic shuttle system similar in some respects to that proposed by
Gorelik et al. has been experimentally realized\cite{12,21}. This situation
has been investigated both in the incoherent and in the quantum case. The
shuttling phenomenon has been observed in Parks et al., experiment\cite{7},
where a C$_{60}$ molecule is oscillating between two leads. In this
experiment, Parks et al., have observed the quantization of current for
various applied voltages, which exhibits step like features within the
current-voltage (I-V) curve. The molecular shuttling between the electrodes in
this experimental setup has been done at low amplitude and high frequency but
provided a key evidence of the involvement of phonon levels in changing the
current properties. This quantized behavior in the conductance characteristics
of NEMS\ devices has been observed in a similar experiment\cite{22}. Another
method to drive the oscillations is to use an external AC electric field
acting on a cantilever (Erbe et al.,)\cite{12}, where the amplitude can be
tuned independently of the source-drain voltage bias. More recently\cite{19},
I-V characteristics of a quantum shuttle device have been observed at room
temperature. These experiments are very close to the original proposal of
Gorelik, et al.

In general, there are two different theoretical formulations that can be used
to study the quantum transport in nanoscopic systems under voltage bias.
Firstly, a generalized quantum master equation approach\cite{23,24,25,26} and
secondly, the nonequilibrium Green's function (NEGF)\cite{27,28} formulation.
The former leads to a simple rate equation, where the coupling between the dot
and the leads is considered as a weak perturbation and the electron-phonon
interaction is also considered very weak. In the latter case one can consider
strong leads to system and electron-phonon coupling. The nonequilibrium
Green's function technique is able to deal with a very broad variety of
physical situations related to shuttling transport\cite{30} at
molecular\cite{31} levels. It can deal with strong non-equilibrium situations
and very small to very large applied bias. In the early seventies, the
nonequilibrium Green's function approach was applied to
mesoscopic\cite{32,33,34,35} transport by Caroli et al., where they were
mainly interested in inelastic transport effects in tunneling through oxide
barriers. This approach was formulated in an elegant way\cite{28,36,37,38,39}
by Mier et al, where they have shown an exact time dependent expression for
the non-equilibrium current through mesoscopic systems. In this model an
interacting and non-interacting mesoscopic system was placed between two large
semi-infinite leads. Furthermore, NEGF has been applied in the study of shot
noise in chain models\cite{40} and disordered junctions\cite{41} while noise
in Coulomb blockade Josephson junctions has been discussed within a phase
correlation theory approach\cite{42}. In the case of an inelastic resonant
tunneling structure, in which strong electron-phonon coupling is often
considered, a very strong source-drain voltage is expected for which coherent
electron transport in molecular devices has been considered by some
workers\cite{43} within the scattering theory approach. Recently, phonon
assisted resonant tunneling conductance has been discussed within the NEGF
technique at zero temperature\cite{44}. To the best of our knowledge, in all
these studies, time-dependent transport properties via a quantum shuttle have
not been discussed so far. The development of time-dependent quantum transport
for the treatment of nonequilibrium system with phononic as well as Fermionic
degrees of freedom has remained a challenge since the
1980's\cite{27,28,36,37,38,39,45}. Generally, time-dependent transport
properties of mesoscopic systems without an oscillating dot have been
reported\cite{46} and, in particular, sudden joining of the leads with a
quantum dot or molecule have been investigated\cite{45,47} for the case of a
noninteracting quantum dot and for a weakly Coulomb interacting molecular
system. Strongly interacting systems in the Kondo regime have been
investigated\cite{48,49}. More recently\cite{50}, the transient effects
occurring in a molecular quantum dot described by an Anderson-Holstein
Hamiltonian has been discussed.

In most of the theoretical work on shuttle based devices since the original
proposal\cite{20}, mechanical degree of freedom has been described
classically/semiclassically or quantum mechanically using the quantum master
or rate equation approach\cite{20,23,24,25,26}. In the original proposal, the
mechanical part was also treated classically, including the damped oscillator,
and assuming an incoherent electron tunneling process. This approach is based
on a perturbation, weak coupling and large applied bias approximations,
whereas the Keldysh nonequilibrium Green's function formulation can treat the
system-leads and electron-phonon coupling with strong interactions\cite{51}
for both small and large applied bias voltage. Moreover, the theories of the
Jauho group\cite{24} and Armour\cite{23} fail to explain the low bias regime.
The transport properties have been described and discussed
semi-classically/classically but need a complete quantum mechanical
description. A theory beyond these cases is required in order to further
refine experiments on quantum shuttle based NEMS\ devices. In the quantum
transport properties of a quantum shuttle; the quantized current can be
determined by the frequency of the quantum mechanical oscillator, the
interplay between the time scales of the electronic and mechanical degrees of
freedom, and the suppression of stochastic tunneling events due to matching of
the Fermionic and oscillator properties. For this reason, a first complete
analytical and qualitative description of the quantum transport properties of
a simple quantum shuttle device is performed and discussed in detail in this work.

In the present work, we shall investigate the time evolution of an oscillating
quantum dot as a reaction to a sudden joining to the leads. We employ the
nonequilibrium Green's function method in order to discuss the transport
properties of a quantum shuttle device. In this system, we include the
time-dependent hopping between the oscillating dot and the leads to enable us
to connect the leads to the dot at a finite time. An advantage of the
time-dependent non-equilibrium Green's function approach considered in this
work is that it eliminates the one major criticism of the Keldysh approach,
which is the lack of any clear initial conditions. We assume strong
dissipation of the oscillating dot, where we consider the average over single
particle properties, implicitly assuming strong damping of the mechanical
oscillator. Simply averaging over a single particle expression is a valid
description of a many particle problem, as long as any excitation caused by
one electron has been dissipated before the next electron arrives. This is the
strong dissipation regime. We describe the electronic state of the dot as a
two-state system (empty/occupied). The electrons in the leads are considered
to be at zero temperature. We consider an arbitrary finite chemical potential
difference between the right and left leads. The transient and the steady
state behavior of the system is considered here in order to explore the
quantum dynamics of the shuttle device as a function of time. This is a fully
quantum mechanical formulation whose basic approximations are very
transparent, as the technique has already been used to study transport
properties in a wide range of mesoscopic systems. In most of the existing
literature very big chemical potential difference is considered while we
include a range from very small to very large. In our calculation inclusion of
the oscillator is not perturbative. As the recent\cite{16} studies are beyond
the perturbation theory, a non-perturbative approach is required beyond the
quantum master equation or linear response. Hence, our work provides an exact
analytical solution to the current--voltage, correlation functions, average
energy, coupling of leads with the system, very small chemical potential
difference and includes both the right and left Fermi level response regimes.
For simplicity, we used the wide-band approximation, where the density of
states in the leads and hence the coupling between the leads and the dot was
taken to be independent of energy. Although the method we used does not rely
on this approximation. This provides a way to perform transient transport
calculations from first principles while retaining the essential physics of
the electronic structure of the dot and the leads. Another advantage of this
method is that it treats the infinitely extended reservoirs in an exact way in
the present system, which may give a better understanding of the essential
features of charge shuttle based NEMS device in a more appropriate quantum
mechanical picture.

\section{Formulation}

The model consists of a moveable quantum dot suspended between the right and
the left semi-infinite leads. This model is a close analogue to the dot
attached to the tip of a cantilever or connected to the leads by some soft
material or embedded into an elastic matrix. In this model the center of mass
of the nanoparticle is confined to a harmonic potential. In the most recent
experimental realizations of a quantum dot shuttle (QDS)\cite{19} at room
temperature, due to its small diameter, the quantum dot has a very small
capacitance and thus has a charging energy that exceeds the thermal energy
k$_{B}$T. Due to Coulomb blockade we consider only one excess electron can
occupy the device. We consider the electronic state of the central quantum dot
as empty or charged (a two level system). The tunneling amplitude of the
electrons depends exponentially on the position of the central island and thus
electrons can tunnel between the leads and the quantum dot. This is due to the
exponentially decreasing/increasing overlapping of the electronic wave functions.

The coupled system is described by a single electronic level of energy
$\epsilon_{0}$ and a nanomechanical oscillator with frequency $\omega$ and
mass $\mu$. An electrostatic force (eE) acts on the mechanical grain when the
quantum dot is charged and gives the electrical influence on the
nanomechanical dynamics of the system. The electric field E is generated by
the voltage drop between the left and the right leads. In our model, though,
it is kept as an external parameter. The quantum dynamics of the
nanomechanical oscillator is represented in terms of position and momentum
operators. In terms of creation and annihilation operators for the
nanomechanical oscillator excitations the Hamiltonian of our simple system
is\cite{20,23,24,25,44,52,53}%

\begin{equation}
H_{0}=H_{dot-ph}+H_{leads}\label{1}%
\end{equation}%
\begin{equation}
H_{\mbox{\scriptsize dot-ph}}=\left[  \epsilon_{0}+\eta(b^{\dagger}+b)\right]
c_{0}^{\dag}c_{0}+H_{ph}\,,\label{2}%
\end{equation}%
\begin{equation}
H_{ph}=\frac{\overset{\wedge}{p}^{2}}{2\mu}%
+{\mathchoice{{\textstyle{\frac12}}}{{\textstyle{\frac12}}}{{\scriptstyle{1/2}}}{{\scriptscriptstyle{1/2}}}}%
\mu\omega\overset{\wedge}{x}^{2}=\hslash\omega(b^{\dagger}%
b+{\mathchoice{{\textstyle{\frac12}}}{{\textstyle{\frac12}}}{{\scriptstyle{1/2}}}{{\scriptscriptstyle{1/2}}}}%
)\,,\label{3}%
\end{equation}
where $\epsilon_{0}$ is the single energy level of electrons on the dot with
$c_{0}^{\dag},c_{0}$ the corresponding creation and annihilation operators,
and an oscillator of frequency $\omega$, mass $\mu$ and $b^{\dagger},b$ are
the raising and lowering operator of the phonons. The parameter $\eta
=\frac{\lambda l}{\sqrt{2}}$ physically represents an effective electric field
in the capacitor seen by the moving dot between the leads, which we shall
refer to as the coupling strength between the moving dot and the electrons on
the dot given as $\lambda l=eEl$, where $e$ is the charge of electron, $E$ is
the strength of electric field and $l=\sqrt{\frac{\hslash}{\mu\omega}}$ is the
zero point amplitude of the oscillator. The remaining elements of the
Hamiltonian are%

\begin{equation}
H_{\mbox{\scriptsize leads}}=\sum_{k}\epsilon_{j}c_{j}^{\dagger}c_{j},
\label{4}%
\end{equation}

\begin{equation}
H_{\mbox{\scriptsize leads-dot}}=\Delta H_{\alpha}=\frac{1}{\sqrt{N}}\sum
_{j}V_{\alpha}(t)\left(  c_{j}^{\dagger}c_{0}+c_{0}^{\dagger}c_{j}\right)  ,
\label{5}%
\end{equation}
where $N$ is the total number of states in the lead, $\alpha=L,R,$ and $j$
represents the channels in one of the leads. We include the time-dependent
hopping $V_{\alpha}(t)$\ to enable us to connect the leads $\alpha$ to the
moving dot at a finite time. An advantage of this approach (joining of leads
to the dot at t=0) is that it eliminates the one major criticism of Keldysh
approach, namely the lack of any clear initial conditions. $V_{\alpha}(t)$ is
written as%

\begin{equation}
V_{\alpha}(t)=V_{0}(t)\mathrm{e}^{\mp\xi\overset{\wedge}{x}}\label{6}%
\end{equation}
where $\xi=\frac{1}{\zeta}$ is the inverse tunneling length ($\zeta$), $-$
stands for $V_{L}(t)=V_{0}(t)\mathrm{e}^{-\xi\overset{\wedge}{x}}$ and + stands for
$V_{R}(t)=V_{0}(t)\mathrm{e}^{+\xi\overset{\wedge}{x}}$. For the time-dependent
dynamics, we shall focus on sudden joining of the leads to the moving dot at
$t=0$, which means $V_{0}(t)=V\theta(t)$, where $\theta(t)$\ is the Heaviside
unit step function. The displacement operator is given as%

\begin{equation}
\overset{\wedge}{x}=\frac{l(b^{\dagger}+b)}{\sqrt{2}}\label{7}%
\end{equation}
The total Hamiltonian of the system is thus $H=H_{0}+\Delta H_{\alpha}$. We
write the eigenvalues and the eigenfunctions of $H_{\mbox{\scriptsize dot-ph}}%
$ as%
\begin{equation}
\epsilon=\epsilon_{0}+\hslash\omega
(n+{\mathchoice{{\textstyle{\frac12}}}{{\textstyle{\frac12}}}{{\scriptstyle{1/2}}}{{\scriptscriptstyle{1/2}}}}%
)-\Delta\label{8}%
\end{equation}%
\begin{equation}
\Psi_{m}(K,x_{0}\neq0)=A_{m}\exp[-{\textstyle\frac{l^{2}K^{2}}{2}}%
]\H_{m}(lK)\exp[-{\mathrm{i}}Kx_{0}]\label{9}%
\end{equation}

\begin{equation}
\Psi_{n}(K,x_{0}=0)=A_{n}\exp[-{\textstyle\frac{l^{2}K^{2}}{2}}]\H_{n}(lK)\,,
\label{10}%
\end{equation}
for the occupied, $x_{0}\not =0$ and unoccupied, $x_{0}=0$, dot respectively,
where $A_{n}=\frac{1}{\sqrt{\sqrt{\pi}2^{n}n!l}},A_{m}=\frac{1}{\sqrt
{\sqrt{\pi}2^{m}m!l}},$ $\Delta=\frac{\lambda^{2}}{2\mu\omega^{2}},$
$x_{0}=\frac{\lambda}{\mu\omega^{2}}$ is the displaced oscillator equilibrium
position due to the coupling to the electrons on the dot, and $\H_{n}(lK)$ are
the usual Hermite polynomials. Here we have used the fact that the harmonic
oscillator eigenfunctions have the same form in both real and Fourier space.

In order to transform between the representations for the occupied and
unoccupied dot we require the matrix with elements $\Phi_{n,m}=\int\Psi
_{n}^{\ast}(K,x_{0}=0)\Psi_{m}(K,x_{0}\neq0)\,{\mathrm{d}}K,$ which may be
simplified\cite{54} as
\begin{eqnarray}
\mathbf{\Phi\equiv}\Phi_{n,m}&=&\sqrt{\frac{2^{|m-n|}\min[n,m]!}{\max[n,m]!}%
}\exp\left(  -\textstyle\frac{1}{4}x^{2}\right)\nonumber\\
&&\times  \left(  \textstyle\frac{1}%
{2}\mathrm{i}x\right)  ^{|m-n|}L_{\min[n,m]}^{|m-n|}\left(  \textstyle\frac
{1}{2}{x^{2}}\right)  \,, \label{11}%
\end{eqnarray}
where $x=\frac{x_{0}}{l}$ is the ratio of equilibrium position of the
displaced oscillator to the zero point amplitude of the oscillator and
$L_{n}^{m-n}(x)$ are the associated Laguerre polynomials. The position of the
resonant level with respect to the chemical potential in the leads is thus
affected by $x$ of the nanomechanical oscillator, which in turn affects the
transport properties of the junction through the device.

\section{The self-energy and the Green's function}

In order to calculate the analytical solutions and to discuss the numerical
results of the transient and steady state quantum dynamics of the
nanomechanical shuttle device, our focus in this section is to derive an
analytical relation for the time-dependent effective self energy and the
corresponding Green's function. The effective self-energy represents the
contribution to the moving dot energy, due to interactions between the
oscillating dot and the leads it is coupled to. In obtaining these results we
use the wide-band approximation only for the simplicity, although the method
we are using does not rely on this approximation. The retarded self-energy of
the oscillating dot due to each lead is given by\cite{28,36,37,38,39,45} (see
Appendix)%
\begin{equation}
\mathbf{\Sigma}^{r}(t,t_{1})=-\frac{\mathrm{i}\Gamma_{\alpha}}{2}\theta(t_{2}%
)\delta(t-t_{1})\mathbf{\Sigma}^{r}(\mp) \label{12}%
\end{equation}
with the matrix $\mathbf{\Sigma}^{r}(\mp)$%
\begin{align}
\mathbf{\Sigma}^{r}(\mp)  &  \equiv\Sigma_{n_{0},n_{0}^{\prime}}^{r}%
(\mp)\nonumber\\
&  =\exp[(\gamma\mp x)^{2}-x^{2}]\sqrt{\frac{2^{|n_{0}^{\prime}-n_{0}|}%
\min[n_{0},n_{0}^{\prime}]!}{\max[n_{0},n_{0}^{\prime}]!}}\nonumber\\
&\phantom{=}\times(\mp\gamma
)^{\left\vert n_{0}^{\prime}-n_{0}\right\vert }L_{\min[n_{0},n_{0}^{\prime}%
]}^{\left\vert n_{0}^{\prime}-n_{0}\right\vert }(-2\gamma^{2})\,,\label{13}
\end{align}
where $\mathbf{\Sigma}^{a}(t,t_{1})=(\mathbf{\Sigma}^{r}(t,t_{1}))^{\ast}$
with $\alpha$\ representing the L or R leads, the dimensionless tunneling
length $\gamma=\frac{\xi l}{\sqrt{2}},$ $x=\frac{x_{0}}{l}$, and the $-,+$
signs stands for the left and the right leads respectively. In the present
representation, the effective self energy is given in terms of a matrix due to
the position dependence of the tunneling matrix elements.

The final result for the lesser self energy matrix may be written as%

\begin{align}
\mathbf{\Sigma}^{<}(t_{1},t_{2}) &  =\mathrm{i}\Gamma_{\alpha}\theta(t_{1})\theta
(t_{2})\times\mathbf{\Sigma}^{<}(\mp)\label{14}\\
&  \times%
%TCIMACRO{\dint _{-\infty}^{\epsilon_{F\alpha}+\frac{1}{2}\hbar\omega}}%
%BeginExpansion
{\displaystyle\int_{-\infty}^{\epsilon_{F\alpha}+\frac{1}{2}\hbar\omega}}
%EndExpansion
\frac{d\varepsilon_{\alpha}}{2\pi}\exp[-\mathrm{i}\varepsilon_{\alpha}(t_{1}%
-t_{2})],\nonumber
\end{align}
where the matrix element of the lesser self-energy is written as%
\begin{equation}
\mathbf{\Sigma}^{<}(\mp)=\mathbf{V}^{<}(\mp)[\mathbf{V}^{<}(\mp)]^{\intercal
},\label{15}%
\end{equation}
where $\mathbf{V}^{<}(\mp)\equiv V_{n}^{<}(\mp)=\frac{\exp[\frac{-x^{2}}%
{2}+(\frac{\gamma\mp x}{2})^{2}]}{\sqrt{n!}}[\mp(\frac{\gamma\pm x}{2})]^{n}$.

This model represent the interplay between two physical time scales of the
system, i.e., the oscillator frequency ($\omega$), and the tunneling rate
($\Gamma_{\alpha}$). The model also shows very interesting interplay between
three physical length scales of the system, i.e., dimensionless tunneling
length ($\gamma$), zero point amplitude ($l$) and the displaced oscillator
equilibrium position to the zero point amplitude of the oscillator ($x$),
which are actually affected by the weak and strong coupling dynamics and small
and large tunneling rate.

We solve Dyson's equation using $H_{\mbox{\scriptsize dot-leads}}$, as a
perturbation in terms of matrix representation. In the presence of the
oscillator, the matrix of the retarded, $\mathbf{G}^{r}(t,t_{1})$ and
advanced, $\mathbf{G}^{a}(t_{2},t^{\prime})$ Green's functions on the dot,
with the phonon states may be written as%
\begin{align}
\mathbf{G}^{r}(t,t_{1}) &  =-\mathrm{i}\theta(t-t_{1})\exp[-\mathrm{i}(\mathbf{M})(t-t_{1}%
)],\text{ \ }t_{1}>0\label{16}\\
&  \equiv G_{m,n_{0}}^{r}(t,t_{1})\nonumber
\end{align}
where $\mathbf{G}^{r}$ and $\mathbf{M}$ are matrices in nanomechanical
oscillator space with indices $m,n_{0}$, and the matrix $\mathbf{M}$\ may be
written as%
\begin{equation}
\mathbf{M}\equiv(E_{m}\delta_{m,n_{0}}+\Sigma_{m,n_{0}}^{r}),\text{
\ \ \ }E_{m}=\epsilon_{0}%
+(m+{\mathchoice{{\textstyle{\frac12}}}{{\textstyle{\frac12}}}{{\scriptstyle{1/2}}}{{\scriptscriptstyle{1/2}}}}%
)\hbar\omega-\Delta,\label{17}%
\end{equation}%
\begin{align}
\mathbf{G}^{a}(t_{2},t^{\prime}) &  =(\mathbf{G}^{r}(t^{\prime},t_{2}))^{\ast
}\nonumber\\
&=+\mathrm{i}\theta(t^{\prime}-t_{2})\exp[-\mathrm{i}(\mathbf{M}^{\ast})(t_{2}-t^{\prime
})],\text{ \ \ }t_{2}>0\nonumber\\
&  \equiv G_{n^{\prime},k}^{a}(t_{2},t^{\prime})\label{18}
\end{align}
The above Eqs. (12), (14), (16), and (18) will be the starting point of our
examination of the time-dependent response of the coupled system. These
functions are the essential ingredients for theoretical consideration of such
diverse problems as low and high voltage, coupling of electron and phonons,
transient and steady state phenomena.

\section{The Density matrix $\rho_{n,n^{\prime}}(t,t)$ and the correlation
function}

The density matrix is related to the lesser Green's function through
$\rho_{n,n^{\prime}}(t,t)=-\mathrm{i}G_{n,n^{\prime}}^{<}(t,t^{\prime})$ at $t^{\prime
}=t$, where the $G_{n,n^{\prime}}^{<}(t,t^{\prime})$\ is the lesser Green's
function\cite{27,28,36,37,38,39,45} on the oscillating dot including all the
contribution from the leads. When $t$ and $t'>0$ the lesser Green's function for the dot in the
presence of the nanomechanical oscillator on the dot, with phonon states in
the representation of the unoccupied dot, may be written as%

\begin{align}
\lefteqn{G_{n,n^{\prime}}^{<}(t,t^{\prime})}\nonumber\\ 
&  =\sum_{m,n_{0},n^{\prime},k}%
%TCIMACRO{\dint _{0}^{t}}%
%BeginExpansion
{\displaystyle\int_{0}^{t}}
%EndExpansion%
%TCIMACRO{\dint _{0}^{t^{\prime}}}%
%BeginExpansion
{\displaystyle\int_{0}^{t^{\prime}}}
%EndExpansion
\mathrm{d}t_{1}\mathrm{d}t_{2}\nonumber\\
&\phantom{=}\times\Phi_{n,m}G_{m,n_{0}}^{r}(t,t_{1})\Sigma_{n_{0},n^{\prime}}^{<}%
(t_{1},t_{2})G_{n^{\prime},k}^{a}(t_{2},t^{\prime})\Phi_{n^{\prime},k}^{\ast
}\nonumber\\
&  \equiv%
%TCIMACRO{\dint _{0}^{t}}%
%BeginExpansion
{\displaystyle\int_{0}^{t}}
%EndExpansion%
%TCIMACRO{\dint _{0}^{t^{\prime}}}%
%BeginExpansion
{\displaystyle\int_{0}^{t^{\prime}}}
%EndExpansion
\mathbf{\Phi G}^{r}(t,t_{1})\mathbf{\Sigma}^{<}(t_{1}%
,t_{2})\mathbf{G}^{a}(t_{2},t^{\prime})\mathbf{\Phi}^{\dagger}
\mathrm{d}t_{1}\mathrm{d}t_{2}\label{19}
\end{align}
whereas for $t$ and $t^{\prime}<0$, the $G_{n,n^{\prime}}^{<}(t,t^{\prime})$
is equal to zero. $G_{n,n^{\prime}}^{<}(t,t^{\prime})$ includes all the
information for the nanomechanical oscillator and electronic leads of the
system, and $m,k,n_{0},n_{1},n,n^{\prime}$ are the oscillator indices. The
lesser self-energy, $\Sigma_{n_{0},n_{1}}^{<}(t_{1},t_{2})$, contains
electronic and oscillator contributions. The electronic contributions are
non-zero only when $t_{1}$ and $t_{2}>0$. The retarded Green's function,
$G_{m,n_{0}}^{r}(t,t_{1})$, is represented as%
\begin{equation}
\mathbf{G}^{r}(t,t_{1})=-\mathrm{i}\theta(t-t_{1})\exp[-\mathrm{i}(\mathbf{M})(t-t_{1})],\text{
}\label{20}%
\end{equation}
Using the eigenvalues and eigenfunctions of $\mathbf{M}$, the retarded Green's
function may be written as%
\begin{equation}
\mathbf{G}^{r}(t,t_{1})=\mathbf{Ug}^{r}(t,t_{1})\mathbf{U}^{\intercal
},\label{21}%
\end{equation}
with eigenvectors $\mathbf{U}$ and its inverse such that $\mathbf{UU}%
^{\intercal}\mathbf{=I}$ (not $\mathbf{UU}^{\dagger}=\mathbf{I}$), and%
\begin{align}
\mathbf{g}^{r}(t,t_{1}) &\equiv g_{i}^{r}(t,t_{1})\nonumber\\
&=-\mathrm{i}\theta(t-t_{1}%
)\exp[-\mathrm{i}(\epsilon_{i}-\mathrm{i}\zeta_{i})(t-t_{1})]\delta_{i,j},\label{22}%
\end{align}
where $\epsilon_{i},\zeta_{i}$ are the real and imaginary parts of the $i$th
eigenvalue of the matrix $\mathbf{M}$. The advanced Green's function,
$G_{n_{1},k}^{a}(t_{2},t)$, is written as%
\begin{equation}
\mathbf{G}^{a}(t_{2},t^{\prime})=+\mathrm{i}\theta(t^{\prime}-t_{2})\exp[-\mathrm{i}(\mathbf{M}%
^{\ast})(t_{2}-t^{\prime})],\text{ }\label{23}%
\end{equation}
In terms of eigenvalues and eigenfunctions, the advanced Green's function is
written as%
\begin{equation}
\mathbf{G}^{a}(t_{2},t^{\prime})=\mathbf{U}^{\ast}\mathbf{g}^{a}%
(t_{2},t^{\prime})\mathbf{U}^{\dagger},\label{24}%
\end{equation}
with eigenvectors $\mathbf{U}^{\ast}$ and its inverse such that $\mathbf{U}%
^{\ast}\mathbf{U}^{\dagger}=\mathbf{I}$, and%
\begin{align}
\mathbf{g}^{a}(t_{2},t^{\prime}) &\equiv g_{i}^{a}(t_{2},t^{\prime}%
)\nonumber\\
&=+\mathrm{i}\theta(t^{\prime}-t_{2})\exp[-\mathrm{i}(\epsilon_{i}+\mathrm{i}\zeta_{i})(t_{2}-t^{\prime
})]\delta_{i,j},\label{25}%
\end{align}
The lesser self-energy, $\mathbf{\Sigma}^{<}(t_{1},t_{2})$, in matrix space of
the oscillator was derived in the previous section.

The lesser Green's function can be calculated by using Eqs. (14,21, \& 24) in
equation (19) at $t=t^{\prime}$ as%
\begin{align}
\lefteqn{G_{n,n^{\prime}}^{<}(t,t)}\nonumber\\  
&  \equiv%
%TCIMACRO{\dint _{0}^{t}}%
%BeginExpansion
{\displaystyle\int_{0}^{t}}
%EndExpansion%
%TCIMACRO{\dint _{0}^{t}}%
%BeginExpansion
{\displaystyle\int_{0}^{t}}
%EndExpansion
\mathrm{d}t_{1}\mathrm{d}t_{2}\mathbf{\Phi Ug}^{r}(t,t_{1})\mathbf{U}^{\intercal}\mathbf{\Sigma
}^{<}(t_{1},t_{2})\mathbf{U}^{\ast}\mathbf{g}^{a}(t_{2},t)\mathbf{U}^{\dag
}\mathbf{\Phi}^{\dag},\nonumber\\
&  =%
%TCIMACRO{\dint _{0}^{t}}%
%BeginExpansion
{\displaystyle\int_{0}^{t}}
%EndExpansion%
%TCIMACRO{\dint _{0}^{t}}%
%BeginExpansion
{\displaystyle\int_{0}^{t}}
%EndExpansion
\mathrm{d}t_{1}\mathrm{d}t_{2}\sum_{g,b,\alpha}[\mathbf{\Phi U}]_{n,g}\mathbf{g}^{r}%
(t,t_{1})\mathbf{U}^{\intercal}\nonumber\\
&\phantom{=}\times\mathbf{\Sigma}^{<}(\mp)\Sigma_{\alpha}%
^{<}(t_{1},t_{2})\mathbf{U}^{\ast}\mathbf{g}^{a}(t_{2},t)[\mathbf{\Phi
U}]_{b,n^{\prime}}^{\dag},\nonumber\\
&  =\sum_{g,b,\alpha}[\mathbf{\Phi U}]_{n,g}[\mathbf{U}^{\intercal
}\mathbf{\Sigma}^{<}(\mp)\mathbf{U}^{\ast}]_{g,b}\chi_{g,b}^{\alpha
}(t)[\mathbf{\Phi U}]_{b,n^{\prime}}^{\dag},\label{26}
\end{align}
where the function $\chi_{g,b}^{\alpha}(t)$\ is written as%

\begin{equation}
\chi_{g,b}^{\alpha}(t)=%
%TCIMACRO{\dint _{0}^{t}}%
%BeginExpansion
{\displaystyle\int_{0}^{t}}
%EndExpansion%
%TCIMACRO{\dint _{0}^{t}}%
%BeginExpansion
{\displaystyle\int_{0}^{t}}
%EndExpansion
\mathrm{d}t_{1}\mathrm{d}t_{2}g_{g}^{r}(t,t_{1})\Sigma_{\alpha}^{<}(t_{1},t_{2})g_{b}^{a}%
(t_{2},t) \label{27}%
\end{equation}
which can be written as%

\begin{align}
\chi_{g,b}^{\alpha}(t)  &  =%
%TCIMACRO{\dint _{0}^{t}}%
%BeginExpansion
{\displaystyle\int_{0}^{t}}
%EndExpansion%
%TCIMACRO{\dint _{0}^{t}}%
%BeginExpansion
{\displaystyle\int_{0}^{t}}
%EndExpansion
\mathrm{d}t_{1}\mathrm{d}t_{2}\exp[-\mathrm{i}(\epsilon_{g}-\mathrm{i}\zeta_{g})(t-t_{1})]\nonumber\\
&\phantom{=}\times\left\{\mathrm{i}\Gamma_{\alpha}%
%TCIMACRO{\dint _{-\infty}^{\epsilon_{F\alpha}+\frac{1}{2}\hbar\omega}}%
%BeginExpansion
{\displaystyle\int_{-\infty}^{\epsilon_{F\alpha}+\frac{1}{2}\hbar\omega}}
%EndExpansion
\frac{d\varepsilon_{\alpha}}{2\pi}\exp[-\mathrm{i}\varepsilon_{\alpha}(t_{1}%
-t_{2})\,\right\}\nonumber\\
&\phantom{=}\times\exp[-\mathrm{i}(\epsilon_{b}+\mathrm{i}\zeta_{b})(t_{2}-t)],\label{28}
\end{align}
Although $\mathbf{G}^{r(a)}(t,t^{\prime})$\ is non-zero for $t<0$, it is never
required due to the way it combines with $\mathbf{\Sigma}_{\alpha}%
^{r,(a),(<)}(t,t^{\prime})$. By carrying out the time integrations, the
resulting expression is written as%

\begin{align}
\lefteqn{\chi_{g,b}^{\alpha}(t)}\nonumber\\  
&=\frac{\mathrm{i}\Gamma_{\alpha}}{2\pi}%
%TCIMACRO{\dint _{-\infty}^{\epsilon_{F\alpha}+\frac{1}{2}\hbar\omega}}%
%BeginExpansion
{\displaystyle\int_{-\infty}^{\epsilon_{F\alpha}+\frac{1}{2}\hbar\omega}}
%EndExpansion
d\varepsilon_{\alpha}\frac{1}{(\varepsilon_{\alpha}-\epsilon_{b}-\mathrm{i}\zeta
_{b})(\varepsilon_{\alpha}-\epsilon_{g}+\mathrm{i}\zeta_{g})}\nonumber\\
&\phantom{=}\times\biggl\{1+\exp[\mathrm{i}(\epsilon_{b}-\epsilon_{g}+\mathrm{i}(\zeta_{b}+\zeta
_{g}))t]\nonumber\\
&\phantom{=\times\{}-\exp[-\mathrm{i}(\varepsilon_{\alpha}-\varepsilon_{b}-\mathrm{i}\zeta_{b}%
)t]\nonumber\\
&\phantom{=\times\{}-\exp[\mathrm{i}(\varepsilon_{\alpha}-\varepsilon_{g}+\mathrm{i}\zeta_{g})t]\biggr\}\label{29}
\end{align}
For all values of $b,g,$and $\alpha$, $\chi_{g,b}^{\alpha}(t=0)$ must be equal
to zero. The integral over the energy in the above equation is carried
out\cite{55}. The final result for the above expression is written as%
\begin{equation}
\chi_{g,b}^{\alpha}(t)=\frac{\mathrm{i}\Gamma_{\alpha}}{2\pi}\{\frac{1}{\varepsilon
_{b}-\varepsilon_{g}+\mathrm{i}(\zeta_{b}+\zeta_{g})}\}[Y_{g,b}^{\alpha}+Z_{g,b}%
^{\alpha}], \label{30}%
\end{equation}
where we have added the contribution from the right and the left leads, which
can be written in terms of$\ \alpha$ as%
\begin{align}
Y_{g,b}^{\alpha}  &  =\left(  1+\exp[\mathrm{i}(\varepsilon_{b}-\varepsilon_{g}%
+\mathrm{i}(\zeta_{b}+\zeta_{g}))t]\right) \nonumber\\
&\phantom{=}\times\biggl\{\frac{1}{2}\frac{\ln[(\epsilon
_{\mathrm{F}\alpha}-\varepsilon_{b})^{2}+\zeta_{b}^{2}]}{\ln[(\epsilon
_{\mathrm{F}\alpha}-\varepsilon_{g})^{2}+\zeta_{g}^{2}]}\nonumber\\
&\phantom{=\times\{}  +\mathrm{i}\left[\tan^{-1}\left(\frac{\varepsilon_{F\alpha}-\varepsilon_{b}}{\zeta_{b}}\right)\right.\nonumber\\
&\phantom{=\times\{+i[}\left.+\tan^{-1}\left(\frac{\varepsilon_{F\alpha}-\varepsilon_{g}}{\zeta_{g}}\right)
+\pi\right]\biggr\},\label{31}
\end{align}
and%
\begin{align}
\lefteqn{Z_{g,b}^{\alpha}}\nonumber\\  
&  =\exp[\mathrm{i}(\varepsilon_{b}-\varepsilon_{g}+\mathrm{i}(\zeta
_{b}+\zeta_{g}))t]\nonumber\\
&\phantom{=}\times\bigl(-\operatorname{Ei}[\mathrm{i}t(\epsilon_{\mathrm{F}\alpha
}-\varepsilon_{b}-\mathrm{i}\zeta_{b})]+\operatorname{Ei}[-\mathrm{i}t(\epsilon_{\mathrm{F}%
\alpha}-\varepsilon_{g}+\mathrm{i}\zeta_{g})]\bigr)\nonumber\\
&\phantom{\times(}  +\operatorname{Ei}[\mathrm{i}t(\epsilon_{\mathrm{F}\alpha}-\varepsilon_{g}%
+\mathrm{i}\zeta_{g})]-\operatorname{Ei}[-\mathrm{i}t(\epsilon_{\mathrm{F}\alpha}-\varepsilon
_{b}-\mathrm{i}\zeta_{b})],\label{32}
\end{align}
with $\epsilon_{\mathrm{F}\alpha}$\ being the right and the left Fermi levels
and $\operatorname{Ei}(x)$ the exponential integral function. Special care is
required in evaluating the $\operatorname{Ei}(x)$ to choose the correct
Riemann sheets in order to make sure that these functions are consistent with
the initial conditions $\chi_{g,b}^{\alpha}(t=0)=0$\ and are continuous
functions of time and chemical potential for each value of b,g, and $\alpha$.

Now using equation (26), the dot population may be written as%
\begin{equation}
\rho(t)=\underset{n}{%
%TCIMACRO{\dsum }%
%BeginExpansion
{\displaystyle\sum}
%EndExpansion
}\rho_{n,n}^{<}(t,t)=\underset{n}{%
%TCIMACRO{\dsum }%
%BeginExpansion
{\displaystyle\sum}
%EndExpansion
}-\mathrm{i}G_{n,n}^{<}(t,t),\label{33}%
\end{equation}
Next we derive a relation for the correlation function which is an interesting
physical quantity to see the shuttling dynamics of the system. With the help
of equation (26) the correlation function Tr$<\overset{\wedge}{p}%
\mathbf{G}^{<}(t,t)>$, where $\overset{\wedge}{p}=\frac{il\mu\omega}{\sqrt{2}%
}(b^{\dagger}-b)$\ is the momentum operator of the oscillator, may be written
as%
\begin{align}
\lefteqn{\text{Tr}\left\langle \frac{il\mu\omega}{\sqrt{2}}(b^{\dagger}-b)\mathbf{G}%
^{<}(t,t)\right\rangle}\nonumber\\ 
&=\underset{n}{%
%TCIMACRO{\dsum }%
%BeginExpansion
{\displaystyle\sum}
%EndExpansion
}\left\langle \frac{il\mu\omega\sqrt{n}}{\sqrt{2}}\{G_{n-1,n}^{<}%
(t,t)-G_{n,n-1}^{<}(t,t)\}\right\rangle ,\label{34}%
\end{align}
which we expect to be finite in case of shuttling as the dot should be
occupied for positive momentum while zero for unoccupied dot. As we note that
the momentum correlation function depends on off-diagonal elements of the
lesser Green's function which can only be finite when nanomechanical
oscillator is in a mixed quantum state rather than in a simple eigenstate.

\section{Time-dependent Current from lead $\alpha$}

The particle current $I_{\alpha}$\ into the interacting region from the lead
is related to the expectation value of the time derivative of the number
operator $N_{\alpha}=\sum_{\alpha j}c_{\alpha j}^{\dagger}c_{\alpha j},$
as\cite{28,36,37,38,39,45,46,47}%

\[
I_{\alpha}=-e\left\langle \frac{\mathrm{d}}{\mathrm{d}t}x\right\rangle
=\frac{\mathrm{i}e}{\hbar}\left\langle [x,H]\right\rangle
\]
and the final result for the current through each of the leads is written as
(See Ref. 51 for detail)\ as%

\begin{align}
I_{\alpha}(t)  &  =\frac{e}{\hbar}%
%TCIMACRO{\dint _{0}^{t}}%
%BeginExpansion
{\displaystyle\int_{0}^{t}}
%EndExpansion
\mathrm{d}t_{1}\nonumber\\
&\phantom{=}\times\text{Tr}\bigl\{\mathbf{G}^{r}(t,t_{1})\mathbf{\Sigma}^{<}(t_{1}%
,t)+\mathbf{G}^{<}(t,t_{1})\mathbf{\Sigma}^{a}(t_{1},t)\nonumber\\
&\phantom{=Tr\{}  -\mathbf{\Sigma}^{r}(t,t_{1})\mathbf{G}^{<}(t_{1},t)-\mathbf{\Sigma}%
^{<}(t,t_{1})\mathbf{G}^{a}(t_{1},t)\bigr\},\nonumber\\
&  =\frac{e}{\hbar}(Z_{1}^{\alpha}+Z_{2}^{\alpha})\label{35}
\end{align}
where $Z_{1}^{\alpha}$ is defined as%

\begin{align}
Z_{1}^{\alpha}&\equiv%
%TCIMACRO{\dint _{0}^{t}}%
%BeginExpansion
{\displaystyle\int_{0}^{t}}
%EndExpansion
\mathrm{d}t_{1}
\text{Tr}\left\{\mathbf{G}^{r}(t,t_{1})\mathbf{\Sigma}^{<}(t_{1}%
,t)-\mathbf{\Sigma}^{<}(t,t_{1})\mathbf{G}^{a}(t_{1},t)\right\}\nonumber\\
&  =%
%TCIMACRO{\dint _{0}^{t}}%
%BeginExpansion
{\displaystyle\int_{0}^{t}}
%EndExpansion
\mathrm{d}t_{1}\text{Tr}\bigl\{\mathbf{Ug}^{r}(t,t_{1})\mathbf{U}^{\intercal}\mathbf{\Sigma
}^{<}(\mp)\Sigma_{\alpha}^{<}(t_{1},t)\nonumber\\
&\phantom{=}\qquad\qquad-\mathbf{\Sigma}^{<}(\mp)\Sigma_{\alpha
}^{<}(t,t_{1})\mathbf{U}^{\ast}\mathbf{g}^{a}(t_{1},t)\mathbf{U}^{\dag
}\bigr\}\nonumber\\
&=\phantom{-}
\underset{g}{\sum}\bigl\{[\mathbf{U}^{\intercal}\mathbf{\Sigma}^{<}%
(\mp)\mathbf{U}]_{g,g}\chi_{g}^{\alpha}(t)\nonumber\\
&\phantom{=}-\underset{g}{\sum}[\mathbf{U}%
^{\dag}\mathbf{\Sigma}^{<}(\mp)\mathbf{U}^{\ast}]_{g,g}\left(  \chi
_{g}^{\alpha}(t)\right)  ^{\ast}\bigr\},\label{36}
\end{align}
where $\chi_{g}^{\alpha}(t)$ is defined as%

\begin{align}
\chi_{g}^{\alpha}(t)  &  =%
%TCIMACRO{\dint _{0}^{t}}%
%BeginExpansion
{\displaystyle\int_{0}^{t}}
%EndExpansion
\mathrm{d}t_{1}g_{g}^{r}(t,t_{1})\Sigma_{\alpha}^{<}(t_{1},t)\nonumber\\
&  =\frac{-\mathrm{i}\theta(t-t_{1})}{2\pi}%
%TCIMACRO{\dint _{0}^{t}}%
%BeginExpansion
{\displaystyle\int_{0}^{t}}
%EndExpansion
\mathrm{d}t_{1}\exp[-\mathrm{i}(\epsilon_{g}-\mathrm{i}\zeta_{g})(t-t_{1})]\nonumber\\
&\phantom{=}\times \mathrm{i}\Gamma_{\alpha}%
%TCIMACRO{\dint _{-\infty}^{\epsilon_{F\alpha}+\frac{1}{2}\hbar\omega}}%
%BeginExpansion
{\displaystyle\int_{-\infty}^{\epsilon_{F\alpha}+\frac{1}{2}\hbar\omega}}
%EndExpansion
d\varepsilon_{\alpha}\exp[-\mathrm{i}\varepsilon_{\alpha}(t_{1}-t)\nonumber\\
&  =\frac{\mathrm{i}\Gamma_{\alpha}}{2\pi}%
%TCIMACRO{\dint _{-\infty}^{\epsilon_{F\alpha}+\frac{1}{2}\hbar\omega}}%
%BeginExpansion
{\displaystyle\int_{-\infty}^{\epsilon_{F\alpha}+\frac{1}{2}\hbar\omega}}
%EndExpansion
d\varepsilon_{\alpha}\frac{1-\exp[-\mathrm{i}(\varepsilon_{\alpha}-\epsilon_{g}%
+\mathrm{i}\zeta_{g})t]}{\varepsilon_{\alpha}-\epsilon_{g}+\mathrm{i}\zeta_{g}}\nonumber\\
&  =\frac{\mathrm{i}\Gamma_{\alpha}}{2\pi}\bigl\{\ln[(\epsilon_{\mathrm{F}\alpha
}-\varepsilon_{g})+\mathrm{i}\zeta_{g}]-\operatorname{Ei}[\mathrm{i}t(\epsilon_{\mathrm{F}%
\alpha}-\varepsilon_{g}+\mathrm{i}\zeta_{g})]\bigr\}\label{37}
\end{align}

Hence, the final result for $Z_{1}^{\alpha}$ is written as%

\begin{align}
Z_{1}^{\alpha}&=\frac{\mathrm{i}\Gamma_{\alpha}}{2\pi}\biggl\{
\underset{g}{\sum}[\mathbf{U}^{\intercal}\mathbf{\Sigma}^{<}(\mp
)\mathbf{U}]_{g,g}\nonumber\\
&\phantom{=}\times\bigl\{\textstyle{\frac{1}{2}}\ln[(\epsilon_{\mathrm{F}\alpha}-\varepsilon
_{g})+\mathrm{i}\zeta_{g}]-\operatorname{Ei}[\mathrm{i}t(\epsilon_{\mathrm{F}\alpha}%
-\varepsilon_{g}+\mathrm{i}\zeta_{g})]\bigr\}\nonumber\\
&\qquad\qquad - \mbox{c.c.}
%\underset{b}{\sum}[\mathbf{U}^{\dag}\mathbf{\Sigma}^{<}(\mp)\mathbf{U}^{\ast
%}]_{b,b}\{\frac{1}{2}\ln[(\epsilon_{\mathrm{F}\alpha}-\varepsilon_{b}%
%)-\mathrm{i}\zeta_{b}]-\operatorname{Ei}[-\mathrm{i}t(\epsilon_{\mathrm{F}\alpha}-\varepsilon
%_{b}-\mathrm{i}\zeta_{b})]\}]
\biggr\}  \label{38}%
\end{align}
and $Z_{2}^{\alpha}$\ is written as%

\begin{align}
\lefteqn{Z_{2}^{\alpha}    \equiv%
%TCIMACRO{\dint }%
%BeginExpansion
{\displaystyle\int}
%EndExpansion
\mathrm{d}t_{1}\text{Tr}\{\mathbf{G}^{<}(t,t_{1})\mathbf{\Sigma}^{a}(t_{1}%
,t)-\mathbf{\Sigma}^{r}(t,t_{1})\mathbf{G}^{<}(t_{1},t)\}}\nonumber\\
&  =%
%TCIMACRO{\dint }%
%BeginExpansion
{\displaystyle\int}
%EndExpansion
\mathrm{d}t_{1}\text{Tr}\biggl\{  \mathbf{G}^{<}(t,t_{1})\mathbf{\Sigma}^{a}(\mp
)\frac{\mathrm{i}\Gamma_{\alpha}}{2}\theta(t_{1})\delta(t-t_{1})\nonumber\\
&\phantom{=}\qquad\qquad+\mathbf{\Sigma}%
^{r}(\mp)\frac{\mathrm{i}\Gamma_{\alpha}}{2}\theta(t_{1})\delta(t-t_{1})\mathbf{G}%
^{<}(t_{1},t)\biggr\} \nonumber\\
&  =\text{Tr}\left\{  \frac{\mathrm{i}\Gamma_{\alpha}}{2}\mathbf{G}^{<}%
(t,t)\mathbf{\Sigma}^{a}(\mp)+\frac{\mathrm{i}\Gamma_{\alpha}}{2}\mathbf{\Sigma}%
^{r}(\mp)\mathbf{G}^{<}(t,t)\right\} \nonumber\\
&  =\mathrm{i}\Gamma_{\alpha}%
%TCIMACRO{\dint _{0}^{t}}%
%BeginExpansion
{\displaystyle\int_{0}^{t}}
%EndExpansion%
%TCIMACRO{\dint _{0}^{t}}%
%BeginExpansion
{\displaystyle\int_{0}^{t}}
%EndExpansion
\mathrm{d}t_{2}\mathrm{d}t_{3}\nonumber\\
&\phantom{=}\text{Tr}\{\mathbf{G}^{r}(t,t_{3})\mathbf{\Sigma}^{<}(\mp
)\Sigma_{\alpha}^{<}(t_{3},t_{2})\mathbf{G}^{a}(t_{2},t)\mathbf{\Sigma}%
^{a}(\mp)\}\nonumber\\
&  =\mathrm{i}\Gamma_{\alpha}%
%TCIMACRO{\dint _{0}^{t}}%
%BeginExpansion
{\displaystyle\int_{0}^{t}}
%EndExpansion%
%TCIMACRO{\dint _{0}^{t}}%
%BeginExpansion
{\displaystyle\int_{0}^{t}}
%EndExpansion
\mathrm{d}t_{2}\mathrm{d}t_{3}\nonumber\\
&\phantom{=}\text{Tr}\{\mathbf{Ug}^{r}(t,t_{3})\mathbf{U}^{\intercal
}\mathbf{\Sigma}^{<}(\mp)\Sigma_{\alpha}^{<}(t_{3},t_{2})\mathbf{U}^{\ast
}\mathbf{g}^{a}(t_{2},t)\mathbf{U}^{\dagger}\mathbf{\Sigma}^{a}(\mp
)\}\nonumber\\
&  =\mathrm{i}\Gamma_{\alpha}\{\underset{g,b,\alpha}{\sum}[[\mathbf{U}^{\dagger
}\mathbf{\Sigma}^{a}(\mp)\mathbf{U}]_{b,g}[\mathbf{U}^{\intercal
}\mathbf{\Sigma}^{<}(\mp)\mathbf{U}^{\ast}]_{g,b}\chi_{g,b}^{\alpha
}(t)\},\label{39}
\end{align}
where $\chi_{g,b}^{\alpha}(t)$ is as in (\ref{30}).

Now using Eqs. (\ref{30}, \ref{38}, \& \ref{39}) in equation (\ref{35}), we arrive at the final result
for the current through lead $\alpha$\ as%

\begin{equation}
I_{\alpha}(t)=\frac{e}{\hbar}(Z_{1}^{\alpha}+Z_{2}^{L}+Z_{2}^{R}), \label{40}%
\end{equation}
where in calculating the left current we need $Z_{1}^{L}$ and both the
contributions $Z_{2}^{L}$ and $Z_{2}^{R}$ and for the right current $Z_{1}%
^{L}$ is replaced by $Z_{1}^{R}$. As before, special care is required in
evaluating the $\ln(x)$ and $\operatorname{Ei}(x)$ to choose the correct
Riemann sheets in order to make sure that these functions are consistent with
the initial conditions $I_{\alpha}(t)=0$\ and are continuous functions of time
and chemical potential.

\section{Average energy}

To calculate the energy transferred from the electrons to the nanomechanical
oscillator, we return to the density matrix $\rho_{n,n}(t,t)$ given in Eq.
(33). We may use the lesser Green's function or density matrix to calculate
the energy transferred to the oscillator as%
\begin{equation}
E_{Ph}=<n>\hslash\omega=\frac{\hslash\omega\underset{n}{\sum}n\rho_{n,n}%
(t,t)}{\underset{n}{\sum}\rho_{n,n}(t,t)} \label{41}%
\end{equation}
where the average evaluated using the diagonal element of the density matrix
on the oscillating quantum dot. Note that the normalization in equation~(42)
is required as the bare density matrix contains both electronic and oscillator
contributions. The trace eliminates the oscillator part, leaving the
electronic part.

\section{Discussion of Results}

The net ($I_{L}(t)-I_{R}(t)$)\ average current through the system, the
correlation function $<\overset{\wedge}{p}G^{<}(t,t)>$, total ($I_{L}%
(t)+I_{R}(t)$)\ average current into the system, and the average energy of an
oscillating dot between the leads are shown graphically as a function of time
(2$\pi/\omega$) for different values of tunneling length, tunneling rate, and
voltage bias. The following parameters were employed: the single energy level
of the dot $\epsilon_{0}=0.5$ and $\epsilon_{\mathrm{F}R}=0.$ These parameters
will remain fixed for all further discussions and have same dimension as
$\hslash\omega$. We are interested in small and large values of tunneling
rates ($\Gamma_{L/R}$) from the leads, different values of the tunneling
length ($\zeta$) between the oscillating dot and the electrons, and of the
left chemical potential $0\leq\epsilon_{\mathrm{F}L}\leq2$. The oscillating
dot induced resonance effects are clearly visible in the numerical results.
The tunneling rate of the electrons between the leads and the dot is
considered to be asymmetric ($\Gamma_{R}=\Gamma_{L}\exp[-\gamma x]$) and we
assume that the leads have constant density of states. The values of the
tunneling rates has been chosen such that when the dot is occupied and the
zero point has moved to x$_{0}$, these values are reversed.

\begin{figure}[htb]
\includegraphics[width=\columnwidth]{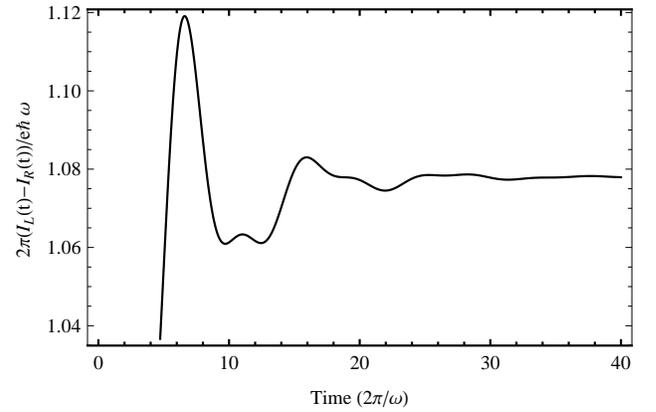}
\caption{\label{fig1} Time-dependent net average current $2\pi(I_{L}(t)-I_{R}%
(t))/e\hbar\omega$ against time (2$\pi/\omega$)\ for fixed values of the
parameters: $\epsilon_{0}=0.5,\epsilon_{FR}=0,\epsilon_{FL}=2,x_{0}%
=0.2l,\hslash\omega=0.63,\Gamma_{L}=\hslash\omega/2\pi,$ $\Gamma_{R}%
=\Gamma_{L}\exp[-\gamma x]$) and $\gamma=0.5$. The electron per cycle behavior
in this figure corresponds to the shuttling mechanism. Units: all the
parameters have same dimension as of $\hslash\omega$.
}
\end{figure}

\begin{figure}[htb]
\includegraphics[width=\columnwidth]{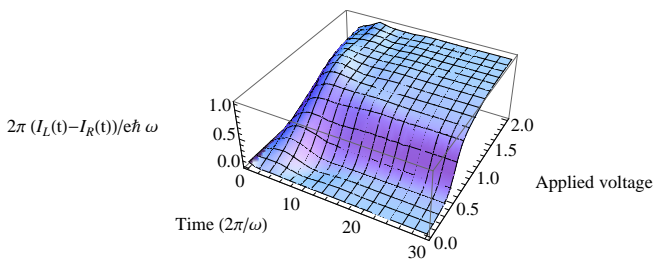}
\caption{\label{fig2} Net average current ($2\pi(I_{L}(t)-I_{R}(t))/e\hbar\omega$)
flowing through the system as a function of both time (2$\pi/\omega$) and of
the left Fermi level for same values of the parameters as in fig.~\ref{fig1}.
}
\end{figure}

\begin{figure}[htb]
\includegraphics[width=\columnwidth]{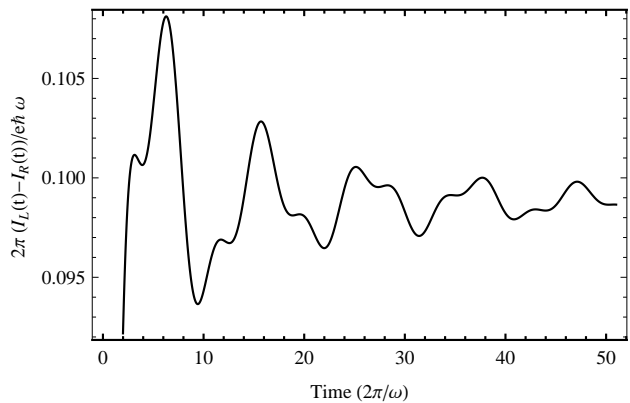}
\caption{\label{fig3}  Net average current ($2\pi(I_{L}(t)-I_{R}(t))/e\hbar\omega$)
flowing through the system as a function of time (2$\pi/\omega$) for fixed
values of the parameters in case of the tunneling mechanism: $\epsilon
_{0}=0.5,\epsilon_{FR}=0,\epsilon_{FL}=2,\hslash\omega=0.63,\Gamma_{L}%
=\Gamma_{R}=0.1\hslash\omega/2\pi,x_{0}=0.1l$ and $\gamma=0.001$. Units: all
the parameters have same dimension as of $\hslash\omega$.
}
\end{figure}

In Fig.~\ref{fig1}~\&~\ref{fig2} we have shown the net average current $2\pi(I_{L}%
(t)-I_{R}(t))/e\hbar\omega$ flowing through the system as a function of time
(2$\pi/\omega$) for \ fixed values of the Fermi level, $\epsilon_{\mathrm{F}%
L}=2$, $\epsilon_{\mathrm{F}R}=0,$ tunneling energy, $\Gamma_{L}=\hbar
\omega/2\pi$, the characteristic energy of the oscillator $\hslash\omega
=0.63$, $\Gamma_{R}=\Gamma_{L}\exp[-\gamma x]$, $x_{0}=0.3l$ and finite
$\gamma=0.5.$ This ($\gamma=0.5$) is the physical condition for shuttling in
the quantum regime: when the zero point amplitude of the nanomechanical
oscillator is half of the tunneling length ($\zeta=\xi^{-1}$). In this figure,
in which the oscillator period is comparable with the left bare tunneling rate
($\Gamma_{L}=\hbar\omega/2\pi$), the average current saturates at one electron
per mechanical cycle (corresponding to average current $2\pi(I_{L}%
(t)-I_{R}(t))/e\hbar\omega=1$) as the electrons are shuttled one by one from
the source to the drain by the oscillating dot. We get the following physical
picture: every time an electron jumps onto the dot when $\Gamma_{L}%
=\hbar\omega/2\pi$, the dot is near the source lead and subject to the
electrostatic force eE that accelerates it towards the drain lead. Energy is
pumped into the nanomechanical oscillator and the dot starts to oscillate
between the leads. When the right tunneling time is high compared to the left
tunnelling time, the oscillator dissipates this energy into the environment
before the next event occurs. This continuously drives the oscillator away
from equilibrium and a stationary state is reached only when the energy pumped
per cycle into the system is dissipated during the same cycle in the
environment. With this setup, the system will be in the shuttle regime. We
note that the frequency of the short period weak oscillations is ($\left\vert
\epsilon_{\mathrm{F}L}-\epsilon_{0}\right\vert $) and these oscillations are
present even in the limit of small but finite $\gamma$. We conclude that this
is a purely electronic process (plasmon oscillations). It is clear from the
figure that in the strong tunneling case, it contains two beating frequencies,
therefore we interpret this as due to a mixture of electronic and mechanical
frequencies. In Fig.~\ref{fig2} the long time behavior of the net average current has
steps as function of the left Fermi energy which become more pronounced with
increasing $\gamma$. In contrast, when both the bare tunneling rates are equal
and smaller than the mechanical frequency, and $\gamma<<1$, the dynamics of
the system is similar to a double barrier resonant tunneling system, since the
dot is static and far from both the leads. The tunneling dynamics of the net
average current is shown in Fig.~\ref{fig3} for fixed values of $\Gamma_{L}=\Gamma
_{R}=0.1\hbar\omega/2\pi,x_{0}=0.1l$ and the oscillator energy $\hbar\omega
=$0.63. This confirms the tunneling behavior of the system oscillating with
beating frequency which is due to mixture of electronic and nanomechanical
degrees of freedom and is consistent with previous studies\cite{51}.

\begin{figure}[htb]
\includegraphics[width=\columnwidth]{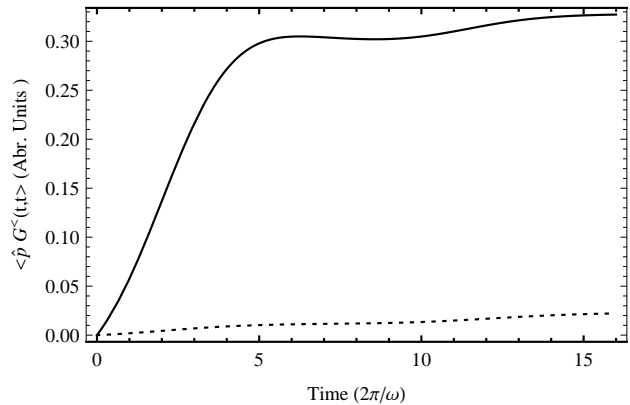}
\caption{\label{fig4} The momentum correlation function as a function of time
(2$\pi/\omega$) showing the shuttling (solid line) and the tunneling (dotted
line) behavior. Parameters for the shuttling are $\epsilon_{0}=0.5,\hslash
\omega=0.63,\Gamma_{L}=\hslash\omega/2\pi,x_{0}=0.2l,$ $\Gamma_{R}=\Gamma
_{L}\exp[-\gamma x]$) and $\gamma=0.5$, and for the tunneling are
$\epsilon_{0}=0.5,\epsilon_{FR}=0,\epsilon_{FL}=2,\hslash\omega=0.63,\Gamma
_{L}=\Gamma_{R}=0.1\hslash\omega/2\pi,x_{0}=0.1l$ and $\gamma=0.001$. All the
parameters have same dimension as $\hslash\omega$.
}
\end{figure}

\begin{figure}[htb]
\includegraphics[width=\columnwidth]{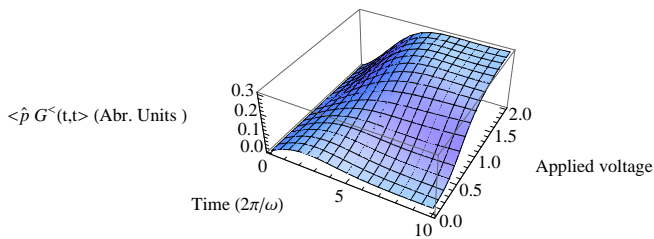}
\caption{\label{fig5}The momentum correlation function as a function of time
(2$\pi/\omega$) and left Fermi energy showing the shuttling behavior.
Parameters for the shuttling are $\epsilon_{0}=0.5,\hslash\omega
=0.63,\Gamma_{L}=\hslash\omega/2\pi,x_{0}=0.2l,$ $\Gamma_{R}=\Gamma_{L}%
\exp[-\gamma x]$) and $\gamma=1$. All the parameters have same dimension as of
$\hslash\omega$.  
}
\end{figure}

The correlation function is shown in Fig.~\ref{fig4}~\&~\ref{fig5}, as a function of time
(2$\pi/\omega$) and left Fermi energy for fixed values of tunneling length
$\zeta=2l$, $x_{0}=0.5l$ , $\hbar\omega=0.63$ and tunneling rates $\Gamma
_{L}=\hbar\omega/2\pi$, $\Gamma_{R}=\Gamma_{L}\exp[-\gamma x]$. We distinguish
between the shuttling and tunneling dynamics of the system for long time for
different tunneling rates and $\gamma$: For shuttling (solid line) $\Gamma
_{L}=\hbar\omega/2\pi$, $\Gamma_{R}=\Gamma_{L}\exp[-\gamma x]$, $\gamma=0.5$,
and for tunneling (dotted line) $\Gamma_{L}=\Gamma_{R}=0.1\hbar\omega/2\pi,$
$x_{0}=0.1l$ , and $\gamma=0.001$. Why have we found this particular type of
structure? We know that, in the ideal shuttling regime, we expect the dot to
be occupied when the momentum is positive and unoccupied when the momentum is
negative. Hence, the correlation function is positive. In contrast, when the
transport is independent of the mechanical oscillator, the correlation
function is zero as $<\overset{\wedge}{p}G^{<}(t,t)>=<\overset{\wedge}%
{p}><G^{<}(t,t)>=0$. We observe that the long time value of the correlation
function has steps as a function of the left Fermi level as shown in
fig.~\ref{fig5}.

\begin{figure}[htb]
\includegraphics[width=\columnwidth]{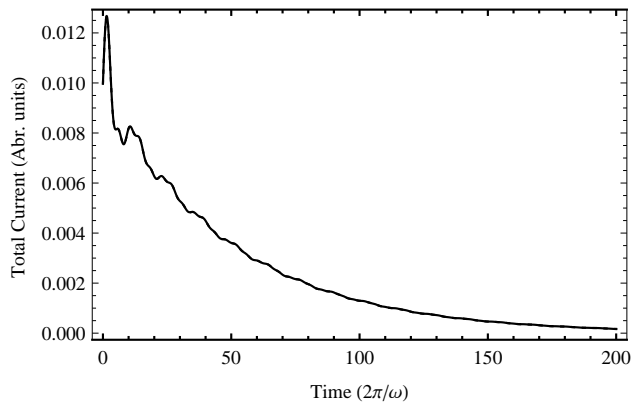}
\caption{\label{fig6} Total average current ($I_{L}(t)+I_{R}(t)$) flowing onto the dot as
a function of time (2$\pi/\omega$) for fixed values of $\epsilon
_{0}=0.5,\epsilon_{FR}=0,\epsilon_{FL}=2,\hslash\omega=0.63,\Gamma_{L}%
=\Gamma_{R}=0.1\hslash\omega/2\pi,x_{0}=0.2l$ and $\gamma=0.1$. All the
parameters have same dimension as $\hslash\omega$. This current (solid line)
is equivalent to the rate of change of dot population $\frac{d}{dt}\rho(t)$
(dashed line) as a function of time for same parameters as of current. In this
figure, solid and dashed lines have same values at all points. Units: all the
parameters have same dimension as of $\hslash\omega$.
}
\end{figure}

Next, we have shown the total average current ($I_{L}(t)+I_{R}(t)$) flowing
onto the oscillating dot in Fig.~\ref{fig6} as a function of time (2$\pi/\omega$) for
fixed values of $\Gamma_{L}=\Gamma_{R}=0.1\hslash\omega/2\pi,$ $\hslash
\omega=0.63,\epsilon_{\mathrm{F}R}=0,$ $\epsilon_{\mathrm{F}L}=2,$ and finite
value of the tunneling length $\gamma=0.1$. This current (solid line) is
equivalent to the rate of change of the dot population (dashed line) for the
same parameters. In this figure, we can not distinguish the solid and the
dashed line. This confirms that our analytical results are consistent with the
equation of continuity, $I_{L}(t)+I_{R}(t)=\frac{d}{dt}\rho(t)$, and hence,
with the conservation laws for all parameters. We interpret the long period
oscillations as a function of time (2$\pi/\omega$) is the mechanical frequency
of the system and short period weak oscillations with the electronic frequency
($\left\vert \epsilon_{\mathrm{F}L}-\epsilon_{0}\right\vert $), which is
consistent with the net average current results as well.

\begin{figure}[htb]
\includegraphics[width=\columnwidth]{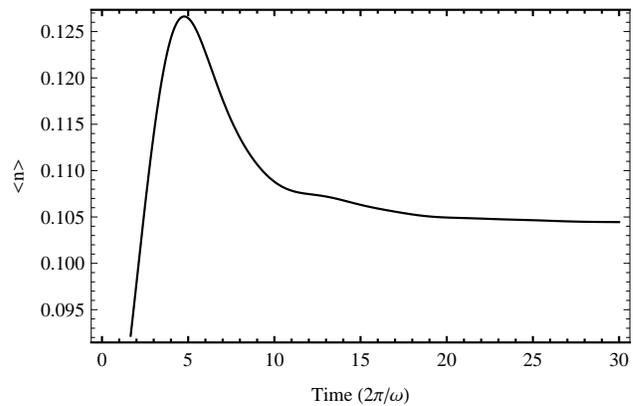}
\caption{\label{fig7}  Average energy transferred to the oscillator as a function of time
(2$\pi/\omega$) for fixed values of $\hslash\omega=0.63,\Gamma_{L}%
=\hslash\omega/2\pi,x_{0}=0.5l,$ $\Gamma_{R}=\Gamma_{L}\exp[-\gamma x]$) and
$\gamma=1$. Units: all the parameters have same dimension as of $\hslash
\omega$.
}
\end{figure}

\begin{figure}[htb]
\includegraphics[width=\columnwidth]{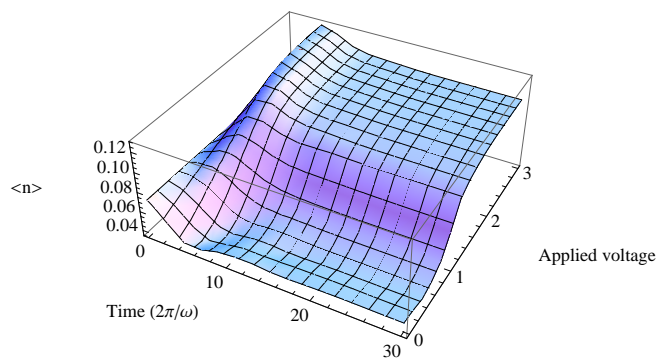}
\caption{\label{fig8} Average energy transferred to the oscillator as a function of time
(2$\pi/\omega$) and left Fermi energy for all the parameters same as in 
fig.~\ref{fig7}.
}
\end{figure}
\begin{figure}[htb]
\includegraphics[width=\columnwidth]{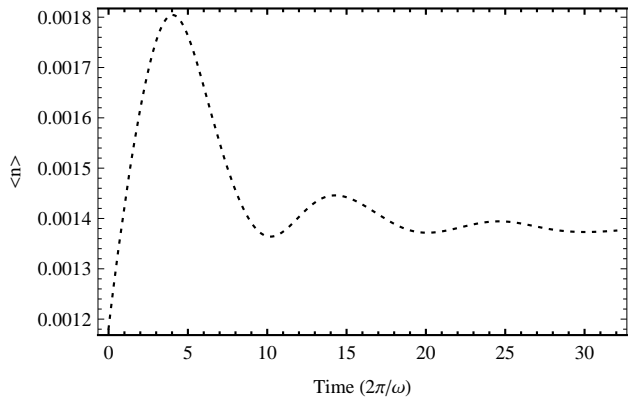}
\caption{\label{fig9} Average energy transferred to the oscillator as a function of time
(2$\pi/\omega$) for fixed values of $\epsilon_{0}=0.5,\epsilon_{FR}%
=0,\epsilon_{FL}=1,\hslash\omega=0.63,\Gamma_{L}=\Gamma_{R}=0.1\hslash
\omega/2\pi,x_{0}=0.5l$ and $\gamma=0.001$. Units: all the parameters have
same dimension as of $\hslash\omega$.
}
\end{figure}

Finally we have shown the average energy of the nanomechanical oscillator as a
function of time (2$\pi/\omega$) and of the left Fermi energy in 
Fig.~\ref{fig7}~\&~\ref{fig8}
for fixed values of tunneling rates $\Gamma_{L}=0.1$, $\Gamma_{R}=\Gamma
_{L}\exp[-\gamma x],\epsilon_{\mathrm{F}R}=0,$ $x_{0}=0.3l$ and tunneling
length $\gamma=0.5$. These parameters corresponds to the shuttling regime of
the system. We found constant average energy of the nanomechanical oscillator
for long time. This constant average energy increases with increasing Fermi
level as shown in fig.~\ref{fig8}. We consider two possible interpretations of this
structure. One is presented in the previous work: that the oscillator is in
any of its pure states and therefore the potential seen by the electrons is
time independent. Secondly, this is an average over several time-dependent
processes with different phases such that any Rabi like features cancel.
Hence, we note that only the second interpretation is consistent with the
correlation function. For the case of the tunneling regime of the system, the
average energy of the nanomechanical oscillator is shown in Fig.~\ref{fig9} as a
function of time (2$\pi/\omega$) for fixed values $\Gamma_{L}=\Gamma
_{R}=0.1\hslash\omega/2\pi,$ $\hslash\omega=0.63,\epsilon_{\mathrm{F}R}=0,$
$\epsilon_{\mathrm{F}L}=2,$ and for very small value of the tunneling length
$l=0.001\zeta.$

In Fig.~\ref{fig7}, \ref{fig8} \& \ref{fig9} it appears that the oscillator is not in its ground state at
t=0. However, if we look at the structure of the lesser self-energy, the
quantity in the middle refers to the state of the oscillator when the dot is
empty and the oscillator is in its ground state. Whereas the whole self-energy
refers to the state of the oscillator when an electron has passed through the
barrier onto the dot. At this point the oscillator is not necessarily in its
ground state. When the electron is in the leads, the oscillator is in its
ground state but when the electron is on the dot two things happen: the
representation of the lesser self-energy is changed and the other is the
effect of the $\exp[\mp\gamma x]$ operator which kicks the oscillator out of
its ground state. This comes back into a saturated state for long time with
finite value of average energy. This could be a mixed or a thermal state. When
taken with the finite correlation function for the same parameters, we
conclude that the long time average energy of the nanomechanical oscillator
corresponds to the mixed state.

\section{Summary}

In this work, we analyzed the time-dependent transport via a quantum shuttle
by using the nonequilibrium Green's function approach without treating the
electron phonon coupling as a perturbation. We have derived an expression for
the full density matrix and discuss it in detail for different values of the
tunneling length and the tunneling rate. Using the full density matrix
calculation we have shown the oscillator momentum charge density correlation
function which distinguishes the shuttling and the tunneling regime of the
system. We derive an expression for the current to see the effects of the
coupling of the electrons to the oscillator on the dot and the tunneling rate
of electrons to resolve the dynamics of the nanomechanical oscillator. We
found a positive correlation function for the nanomechanical oscillator in the
shuttling regime and zero in the case of the tunneling regime. The shuttling
regime will occur when the electron left jump rate introduces continued kicks
on the island. This happens when the left jump rate is close to the
oscillation frequency of the island $\Gamma_{L}=\hbar\omega/2\pi$. By setting
an appropriate ejection rate ($\Gamma_{R}=\Gamma_{L}\exp[-\gamma x]$) and
$l=0.5\zeta$, there exists a condition where the island will keep oscillating
between the leads. In contrast, in the limit of equal bare tunneling rates,
smaller than the mechanical frequency, and when $\gamma<<1$, the dynamics of
the system is similar to a double barrier resonant tunneling system, where we
found periodic oscillations as a function of time (2$\pi/\omega$), which
corresponds to the mechanical frequency of the system. Furthermore, we discuss
the average energy transferred to oscillator as a function of time. We have
found a mixed or a thermal state shown in the average energy of the
nanomechanical oscillator for long time which is consistent with the finite
correlation function for the same parameters. These results suggest further
experiments for NEMS to go beyond the classical dynamics of a shuttle device.

\appendix

\section{\label{App.A}}

In this appendix we derive an analytical relation for the time-dependent
self-energy. The effective self-energy represents the contribution to the
moving dot energy, due to interactions between the oscillating dot and the
leads it is coupled to. The retarded self-energy of the oscillating dot due to
lead $\alpha$\ is given by%
\begin{equation}
\Sigma_{n,n^{\prime}}^{r}(t,t_{1})=\phi_{n,m}(t)g_{\alpha,\alpha
}^{r}(t,t_{1})\phi_{n^{\prime},m}^{\ast}(t_{1}),\label{A1}%
\end{equation}
where the uncoupled Green's function $g_{\alpha,\alpha}^{r}(t,t_{1})$ in the
leads is%
\begin{align}
\lefteqn{g_{\alpha,\alpha}^{r}(t,t_{1}) =\frac{1}{N}\sum_{j}g_{\alpha,j}^{r}%
(t,t_{1})=\overset{+\infty}{\underset{-\infty}{%
%TCIMACRO{\dint }%
%BeginExpansion
{\displaystyle\int}
%EndExpansion
}}n_{\alpha}d\varepsilon_{\alpha}g_{\alpha}^{r}(t,t_{1})}\label{A2}\\
&  =-\mathrm{i}\theta(t-t_{1})\overset{+\infty}{\underset{-\infty}{%
%TCIMACRO{\dint }%
%BeginExpansion
{\displaystyle\int}
%EndExpansion
}}d\varepsilon_{\alpha}n_{\alpha}\exp[-\mathrm{i}(\varepsilon_{\alpha}%
+(m+{\mathchoice{{\textstyle{\frac12}}}{{\textstyle{\frac12}}}{{\scriptstyle{1/2}}}{{\scriptscriptstyle{1/2}}}}%
)\hbar\omega)(t-t_{1})],\nonumber
\end{align}
and $\underset{j}{%
%TCIMACRO{\dsum }%
%BeginExpansion
{\displaystyle\sum}
%EndExpansion
}\Rightarrow\overset{+\infty}{\underset{-\infty}{%
%TCIMACRO{\dint }%
%BeginExpansion
{\displaystyle\int}
%EndExpansion
}}Nn_{\alpha}d\varepsilon_{\alpha},$\ where $j$ stands for every channel in
each lead, $n_{\alpha}$ is the density of states in lead $\alpha$,
$g_{\alpha,j}^{r}(t,t_{1})$\ is the uncoupled Green's function for every
channel in the leads. The tunneling matrix $\phi_{n,m}(t)$\ is written as%
\begin{align}
\phi_{n,m}(t)  &=\left<V_{0}(t)\mathrm{e}^{\mp\xi\overset{\wedge}{x}}\right>\nonumber\\
&  = \biggl\{
%TCIMACRO{\dint }%
%BeginExpansion
{\displaystyle\int}
%EndExpansion
A_{n}^{\ast}\mathrm{e}^{-\frac{(x-x_{0})^{2}}{2l^{2}}}\H_{n}\left(\frac{x-x_{0}}%
{l}\right)\nonumber\\
&\phantom{=}\quad\quad\quad V_{0}(t)\mathrm{e}^{\mp\xi\overset{\wedge}{x}}\nonumber\\
&\phantom{=}\quad\quad\quad\quad A_{m}\mathrm{e}^{-\frac{(x)^{2}}{2l^{2}}}%
\H_{m}\left(\frac{x}{l}\right)\biggr\}\mathrm{d}x,\label{A3}
\end{align}
where $x$ is the coordinate of the oscillator, $\xi$\ being the inverse
tunneling length.

Using the expressions of the tunneling matrix and the uncoupled retarded
Green's function in equation (A1), the retarded self-energy may be written as%
\begin{widetext}
\begin{align}
\Sigma_{n,n^{\prime}}^{r}(t,t_{1})
 &  =\,\sum_{m}-\mathrm{i}\theta
(t-t_{1})\overset{+\infty}{\underset{-\infty}{%
%TCIMACRO{\dint }%
%BeginExpansion
{\displaystyle\int}
%EndExpansion
}}Nd\varepsilon_{\alpha}\frac{\left\vert V\right\vert ^{2}\theta
(t)\theta(t_{1})}{N}n_{\alpha}\exp[-\mathrm{i}(\varepsilon_{\alpha}%
+(m+{\mathchoice{{\textstyle{\frac12}}}{{\textstyle{\frac12}}}{{\scriptstyle{1/2}}}{{\scriptscriptstyle{1/2}}}}%
)\hbar\omega)(t-t_{1})]\nonumber\\
%TCIMACRO{\dint }%
%BeginExpansion
&\phantom{=}\times{\displaystyle\int}
%EndExpansion%
%TCIMACRO{\dint }%
\mathrm{d}x A_{n}\mathrm{e}^{-\frac{(x-x_{0})^{2}}{2l^{2}}}\H_{n}\left(\frac
{x-x_{0}}{l}\right)V_{0}\mathrm{e}^{\mp\xi\overset{\wedge}{x}}A_{m}^{\ast}\mathrm{e}^{-\frac{(x)^{2}%
}{2l^{2}}}
\H_{m}\left(\frac{x}{l}\right)\nonumber\\
%BeginExpansion
&\phantom{=}\times{\displaystyle\int}\mathrm{d}x^{\prime}
%EndExpansion
A_{n^{\prime}}^{\ast}\mathrm{e}^{-\frac{(x^{\prime
}-x_{0})^{2}}{2l^{2}}}\H_{n^{\prime}}\left(\frac{x^{\prime}-x_{0}}{l}%
\right)V_{0}\mathrm{e}^{\mp\xi\overset{\wedge}{x}^{\prime}}A_{m}\mathrm{e}^{-\frac{(x^{\prime})^{2}%
}{2l^{2}}}\H_{m}\left(\frac{x^{\prime}}{l}\right),\label{A4}\\
%\Sigma_{n,n^{\prime}}^{r}(t,t_{1}) 
&  =\sum_{m}-\mathrm{i}\theta(t-t_{1}%
)\left\vert V\right\vert ^{2}\theta(t)\theta(t_{1})n_{\alpha}
\overset{+\infty}{\underset{-\infty}{%
%TCIMACRO{\dint }%
%BeginExpansion
{\displaystyle\int}
%EndExpansion
}}\exp[-\mathrm{i}(\varepsilon_{\alpha}%
+(m+{\mathchoice{{\textstyle{\frac12}}}{{\textstyle{\frac12}}}{{\scriptstyle{1/2}}}{{\scriptscriptstyle{1/2}}}}%
)\hbar\omega)(t-t_{1})]d\varepsilon_{\alpha}\nonumber\\
%TCIMACRO{\dint }%
&\phantom{=}\times
%BeginExpansion
{\displaystyle\int}\mathrm{d}x
%EndExpansion%
%TCIMACRO{\dint }%
A_{n}\exp\left(-\frac{(x-x_{0})^{2}}{2l^{2}}\right)\H_{n}\left(\frac
{x-x_{0}}{l}\right)V_{0}\mathrm{e}^{\mp\xi\overset{\wedge}{x}}A_{m}^{\ast}
\exp\left(-\frac{(x)^{2}%
}{2l^{2}}\right) \H_{m}\left(\frac{x}{l}\right)\nonumber\\
& \phantom{=} \times
%BeginExpansion
{\displaystyle\int}
%EndExpansion
\mathrm{d}x^{\prime}A_{n^{\prime}}^{\ast}\exp\left(-\frac{(x^{\prime
}-x_{0})^{2}}{2l^{2}}\right)\H_{n^{\prime}}\left(\frac{x^{\prime}-x_{0}}{l}%
\right)V_{0}\mathrm{e}^{\mp\xi\overset{\wedge}{x}^{\prime}}A_{m}
\exp\left(-\frac{(x^{\prime})^{2}%
}{2l^{2}}\right)\H_{m}\left(\frac{x^{\prime}}{l}\right),\label{A5}\\
&  =-\mathrm{i}n_{\alpha}\left\vert V\right\vert ^{2}\theta(t)\theta(t_{1}%
)\theta(t-t_{1})2\pi\delta(t-t_{1})\nonumber\\
&\phantom{=}\times\sum_{m}%
%TCIMACRO{\dint }%
%BeginExpansion
{\displaystyle\int}
%EndExpansion%
\mathrm{d}x
%TCIMACRO{\dint }%
A_{n}\exp\left(-\frac{(x-x_{0})^{2}}{2l^{2}}\right)
\H_{n}\left(\frac{x-x_{0}}{l}\right)
V_{0}\mathrm{e}^{\mp\xi\overset{\wedge}{x}}A_{m}^{\ast}\exp\left(-\frac{(x)^{2}%
}{2l^{2}}\right)\H_{m}\left(\frac{x}{l}\right)\nonumber\\
&\phantom{=\sum_n}\times
%BeginExpansion
{\displaystyle\int}
%EndExpansion
\mathrm{d}x^{\prime}A_{n^{\prime}}^{\ast}\exp\left(-\frac{(x^{\prime
}-x_{0})^{2}}{2l^{2}}\right)\H_{n^{\prime}}\left(\frac{x^{\prime}-x_{0}}{l}%
\right)V_{0}\mathrm{e}^{\mp\xi x^{\prime}}A_{m}\exp\left(-\frac{(x^{\prime})^{2}}{2l^{2}}\right)%
\H_{m}\left(\frac{x^{\prime}}{l}\right),\label{A6}
\end{align}
where $\Gamma_{\alpha}$=$4\pi\left\vert V\right\vert ^{2}n_{\alpha}$. Using
the completeness identity, 
\[\underset{m}{\sum}A_{m}^{\ast}\exp\left(-\frac{(x)^{2}%
}{2l^{2}}\right)\H_{m}\left(\frac{x}{l}\right)
A_{m}\exp\left(-\frac{(x^{\prime})^{2}}{2l^{2}}\right)%
\H_{m}\left(\frac{x^{\prime}}{l}\right)=\delta(x-x^{\prime})
\] 
the above equation can be simplified as%
\begin{align}
\Sigma_{n,n^{\prime}}^{r}(t,t_{1}) &  =-\frac{\mathrm{i}\Gamma_{\alpha}}%
{2}\theta(t_{1})\delta(t-t_{1})\label{A7}\\
&  \times%
%TCIMACRO{\dint }%
%BeginExpansion
{\displaystyle\int}
%EndExpansion
\mathrm{d}x^{\prime}A_{n}\exp\left(-\frac{(x^{\prime}-x_{0})^{2}}{l^{2}}\right)
\H_{n}\left(\frac{x^{\prime}-x_{0}}{l}\right)V_{0}\mathrm{e}^{\mp2\xi\overset{\wedge}{x}^{\prime}%
}A_{n^{\prime}}^{\ast}\exp\left(-\frac{(x^{\prime}-x_{0})^{2}}{2l^{2}}\right)%
\H_{n^{\prime}}\left(\frac{x^{\prime}-x_{0}}{l}\right),\nonumber
\end{align}
\end{widetext}
After integrating the above expression, we arrive at the following final
result for the retarded self-energy\cite{54},%
\begin{align}
\Sigma_{n,n^{\prime}}^{r}(t,t_{1}) &  =-\frac{\mathrm{i}\Gamma_{\alpha}}%
{2}\theta(t_{1})\delta(t-t_{1})\mathbf{\Sigma}^{r}(\mp),\label{A8}\\
&  =\Sigma_{\alpha}^{r}(t,t_{1})\mathbf{\Sigma}^{r}(\mp)\equiv\mathbf{\Sigma
}^{r}(t,t_{1})\nonumber
\end{align}
with%
\begin{equation}
\mathbf{\Sigma}^{r}(\mp)\equiv\psi_{n,n^{\prime}}^{\mp}\exp[(\gamma\mp
x)^{2}-x^{2}]\label{A9}%
\end{equation}
where $\Sigma_{\alpha}^{a}(t,t_{1})=(\Sigma_{\alpha}^{r}(t,t_{1}))^{\ast}$
with $\alpha$\ represents the L or R leads, and the matrices $\psi
_{n,n^{\prime}}$\ is given as%
\begin{equation}
\psi_{n,n^{\prime}}^{\mp}=\sqrt{\frac{2^{|n^{\prime}-n|}%
\min[n,n^{\prime}]!}{\max[n,n^{\prime}]!}}(\mp\gamma
)^{\left\vert n^{\prime}-n\right\vert }{\mathop{\mathrm{L}}\nolimits}_{\min[n,n^{\prime}%
]}^{\left\vert n^{\prime}-n\right\vert }(-2\gamma^{2}),\ \label{A10}%
\end{equation}
where the dimensionless tunneling length $\gamma=\frac{\xi l}{\sqrt{2}},$
$x=\frac{x_{0}}{l}$, and the $-,+$ signs stands for the left and the right
leads respectively.

Similarly, the lesser self energy may be calculated as%
\begin{equation}
\Sigma_{n,n^{\prime}}^{<}(t_{1},t_{2})=\phi_{n,m}(t_{1})g_{\alpha,\alpha
}^{<}(t_{1},t_{2})\phi_{n^{\prime},m}^{\ast}(t_{2}),\label{A11}%
\end{equation}
where $g_{\alpha,\alpha}^{<}(t_{1},t_{2})$ is the uncoupled Green's function
in the leads and is given as%
\begin{align}
g_{\alpha,\alpha}^{<}(t_{1},t_{2}) &  =\frac{1}{N}\sum_{j}g_{\alpha,j}%
^{<}(t_{1},t_{2})=\overset{+\infty}{\underset{-\infty}{%
%TCIMACRO{\dint }%
%BeginExpansion
{\displaystyle\int}
%EndExpansion
}}n_{\alpha}d\varepsilon_{\alpha}g_{\alpha}^{<}(t_{1},t_{2})\label{A12}\\
&  =\overset{+\infty}{\underset{-\infty}{%
%TCIMACRO{\dint }%
%BeginExpansion
{\displaystyle\int}
%EndExpansion
}}2\mathrm{i}n_{\alpha}f(\varepsilon_{\alpha})B_{m}d\varepsilon_{\alpha}\exp
[-\mathrm{i}\varepsilon_{\alpha}(t_{1}-t_{2})],\nonumber
\end{align}
We are going to consider that temperature is zero in the leads. So, we assume,
when the oscillator appear in the leads it will always be in its ground state
such that $m=0$ ($B_{m}=\delta_{m,0}$) and the Fermi distribution function can
be replaced with the theta function which in turn limits the integral. With
the help of these approximations, and using the expressions for the tunneling
matrix and the uncoupled retarded Green's function into equation (A11), the
lesser self-energy may be written as%
\begin{widetext}
\begin{align}
\Sigma_{n,n^{\prime}}^{<}(t_{1},t_{2}) &  =\,%
%TCIMACRO{\dint _{-\infty}^{\epsilon_{F\alpha}+\frac{1}{2}\hbar\omega}}%
%BeginExpansion
{\displaystyle\int_{-\infty}^{\epsilon_{F\alpha}+\frac{1}{2}\hbar\omega}}
%EndExpansion
N2\mathrm{i}n_{\alpha}d\varepsilon_{\alpha}\frac{\left\vert V\right\vert ^{2}}{N}%
\theta(t_{1})\theta(t_{2})\exp[-\mathrm{i}\varepsilon_{\alpha}(t_{1}-t_{2}%
)]\nonumber\\
&\phantom{=}  \times%
%TCIMACRO{\dint }%
%BeginExpansion
{\displaystyle\int}
%EndExpansion
\mathrm{d}x A_{n}\exp\left(-\frac{(x-x_{0})^{2}}{2l^{2}}\right)
\H_{n}\left(\frac{x-x_{0}}{l}\right)
V_{0}\mathrm{e}^{\mp\xi\overset{\wedge}{x}}A_{0}^{\ast}
\exp\left(-\frac{(x)^{2}}{2l^{2}}\right)\nonumber\\
& \phantom{=} \times%
%TCIMACRO{\dint }%
%BeginExpansion
{\displaystyle\int}
%EndExpansion
\mathrm{d}x^{\prime}A_{n^{\prime}}^{\ast}
\exp\left(-\frac{(x^{\prime}-x_{0})^{2}}{2l^{2}}\right)
\H_{n^{\prime}}\left(\frac{x^{\prime}x_{0}}{l}\right)
V_{0}\mathrm{e}^{\mp\xi\overset{\wedge}{x}^{\prime}}%
A_{0}\exp\left(-\frac{(x^{\prime})^{2}}{2l^{2}}\right),\label{A13}\\
%\Sigma_{n,n^{\prime}}^{<}(t_1,t_2) 
&  =2\mathrm{i}n_{\alpha}\left\vert V\right\vert ^{2}%
\theta(t_{1})\theta(t_{2})\,%
%TCIMACRO{\dint _{-\infty}^{\epsilon_{F\alpha}+\frac{1}{2}\hbar\omega}}%
%BeginExpansion
{\displaystyle\int_{-\infty}^{\epsilon_{F\alpha}+\frac{1}{2}\hbar\omega}}
%EndExpansion
d\varepsilon_{\alpha}\exp[-\mathrm{i}\varepsilon_{\alpha}(t_{1}-t_{2})]\nonumber\\
&\phantom{=} \times%
%TCIMACRO{\dint }%
%BeginExpansion
{\displaystyle\int}
%EndExpansion
\mathrm{d}x A_{n}\exp\left(-\frac{(x-x_{0})^{2}}{2l^{2}}\right)
\H_{n}\left(\frac{x-x_{0}}{l}\right)
V_{0}\mathrm{e}^{\mp\xi\overset{\wedge}{x}}A_{0}^{\ast}
\exp\left(-\frac{(x)^{2}}{2l^{2}}\right)%
\nonumber\\
&\phantom{=}  \times%
%TCIMACRO{\dint }%
%BeginExpansion
{\displaystyle\int}
%EndExpansion
\mathrm{d}x^{\prime} A_{n^{\prime}}^{\ast}\exp\left(-\frac{(x^{\prime}-x_{0})^{2}}{2l^{2}}\right)
\H_{n^{\prime}}\left(\frac{x^{\prime}-x_{0}}{l}\right)
V_{0}\mathrm{e}^{\mp\xi\overset{\wedge}{x}^{\prime}}%
A_{0}\exp\left(-\frac{(x^{\prime})^{2}}{2l^{2}}\right)\,,\label{A14}\\
&  =\mathrm{i}\Gamma_{\alpha}\theta(t_{1})\theta(t_{2})%
%TCIMACRO{\dint _{-\infty}^{\epsilon_{F\alpha}+\frac{1}{2}\hbar\omega}}%
%BeginExpansion
{\displaystyle\int_{-\infty}^{\epsilon_{F\alpha}+\frac{1}{2}\hbar\omega}}
%EndExpansion
\frac{d\varepsilon_{\alpha}}{2\pi}\exp[-\mathrm{i}\varepsilon_{\alpha}(t_{1}%
-t_{2})]\nonumber\\
&\phantom{=} \times%
%TCIMACRO{\dint }%
%BeginExpansion
{\displaystyle\int}
%EndExpansion
\mathrm{d}x A_{n}\exp\left(-\frac{(x-x_{0})^{2}}{2l^{2}}\right)
\H_{n}\left(\frac{x-x_{0}}{l}\right)
V_{0}\mathrm{e}^{\mp\xi\overset{\wedge}{x}}A_{0}^{\ast}
\exp\left(-\frac{(x)^{2}}{2l^{2}}\right)\nonumber\\
&\phantom{=}\times%
%TCIMACRO{\dint }%
%BeginExpansion
{\displaystyle\int}
%EndExpansion
\mathrm{d}x^{\prime} A_{n^{\prime}}^{\ast}
\exp\left(-\frac{(x^{\prime}-x_{0})^{2}}{2l^{2}}\right)
\H_{n^{\prime}}\left(\frac{x^{\prime}-x_{0}}{l}\right)
V_{0}\mathrm{e}^{\mp\xi\overset{\wedge}{x}^{\prime}}%
A_{0}\exp\left(-\frac{(x^{\prime})^{2}}{2l^{2}}\right),\label{A15}
\end{align}
where $\Gamma_{\alpha}$=$4\pi\left\vert V_{0,\alpha}\right\vert ^{2}n_{\alpha
}$, with $\alpha$\ representing the L or R leads, and the above equation can
be rewritten as%
\begin{align}
\Sigma_{n,n^{\prime}}^{<}(t_{1},t_{2}) &  =\mathrm{i}\Gamma_{\alpha}\theta(t_{1}%
)\theta(t_{2})\,%
%TCIMACRO{\dint _{-\infty}^{\epsilon_{F\alpha}+\frac{1}{2}\hbar\omega}}%
%BeginExpansion
{\displaystyle\int_{-\infty}^{\epsilon_{F\alpha}+\frac{1}{2}\hbar\omega}}
%EndExpansion
\frac{d\varepsilon_{\alpha}}{2\pi}\exp[-\mathrm{i}\varepsilon_{\alpha}(t_{1}%
-t_{2})]\nonumber\\
&\phantom{=}\times%
%TCIMACRO{\dint }%
%BeginExpansion
{\displaystyle\int}
%EndExpansion
\mathrm{d}x^{\prime} A_{n}\exp\left(-\frac{(x^{\prime}-x_{0})^{2}}{l^{2}}\right)
\H_{n}\left(\frac{x^{\prime}-x_{0}}{l}\right)
V_{0}\mathrm{e}^{\mp\xi\overset{\wedge}{x}^{\prime}}%
A_{0}^{\ast}\exp\left(-\frac{x^{\prime2}}{2l^{2}}\right)\nonumber\\
& \phantom{=} \times%
%TCIMACRO{\dint }%
%BeginExpansion
{\displaystyle\int}
%EndExpansion
\mathrm{d}x A_{n^{\prime}}\exp\left(-\frac{(x-x_{0})^{2}}{2l^{2}}\right)
\H_{n^{\prime}}\left(\frac{x-x_{0}}{l}\right)
V_{0}\mathrm{e}^{\mp\xi\overset{\wedge}{x}}A_{0}^{\ast}
\exp\left(-\frac{(x)^{2}}{2l^{2}}\right),\label{A16}
\end{align}
Which can be simplified\cite{54} by carrying out the integrals and the final
result is written as%
\begin{align}
\Sigma_{n,n^{\prime}}^{<}(t_{1},t_{2}) &  =\mathrm{i}\Gamma_{\alpha}\theta(t_{1}%
)\theta(t_{2})\frac{\exp[\frac{-x^{2}}{2}+(\frac{\gamma\mp x}{2})^{2}]}%
{\sqrt{n!}}[\mp(\frac{\gamma\pm x}{2})]^{n}\label{A17}\\
&  \times\frac{\exp[\frac{-x^{2}}{2}+(\frac{\gamma\mp x}{2})^{2}]}{\sqrt
{n^{\prime}!}}[\mp(\frac{\gamma\pm x}{2})]^{n^{\prime}}\,%
%TCIMACRO{\dint _{-\infty}^{\epsilon_{F\alpha}+\frac{1}{2}\hbar\omega}}%
%BeginExpansion
{\displaystyle\int_{-\infty}^{\epsilon_{F\alpha}+\frac{1}{2}\hbar\omega}}
%EndExpansion
\frac{d\varepsilon_{\alpha}}{2\pi}\exp[-\mathrm{i}\varepsilon_{\alpha}(t_{1}%
-t_{2})],\nonumber\\
&  =\mathbf{\Sigma}^{<}(\mp)\Sigma_{\alpha}^{<}(t_{1},t_{2})\equiv
\mathbf{\Sigma}^{<}(t_{1},t_{2})\nonumber
\end{align}
where $\gamma=\frac{l\xi}{\sqrt{2}}$, $x=\frac{x_{0}}{l}$, and the $-,+$ signs
stands for the left and the right leads respectively.
\end{widetext}
ACKNOWLEDGMENT

M.Tahir would like to acknowledge the support of the Pakistan Higher Education
Commission (HEC).

$^{\ast}$Permanent address: Department of Physics, University of Sargodha,
Sargodha, Pakistan; m.tahir@uos.edu.pk, m.tahir06@imperial.ac.uk

\end{document}